# Robustness of Skyrmion-Mediated Magnetization Switching in Perpendicular Magnetic Tunnel Junctions (p-MTJs) to Material Disorder and Elevated Temperature

Prosenjit Das,[1] Md Mahadi Rajib,[1] and Jayasimha Atulasimha[1,2*]

[1]Department of Mechanical and Nuclear Engineering, Virginia Commonwealth University, Richmond, Virginia, USA

[2]Department of Electrical and Computer Engineering, Virginia Commonwealth University, Richmond, Virginia, USA

*Corresponding author: jatulasimha@vcu.edu

## Abstract

Transient-skyrmion mediated voltage-controlled magnetic anisotropy (VCMA) switching is a fast and energy efficient way to magnetization reversal, but its reliability under realistic material disorder and thermal fluctuations remain unclear. In this work, we use stochastic micromagnetic simulations of a 100 nm perpendicular magnetic tunnel junction to study switching reliability under grain-scale variations in Dzyaloshinskii–Moriya interaction, perpendicular magnetic anisotropy, easy-axis orientation, and intergrain exchange coupling. We find that switching remains robust under moderate disorder, with 100% switching observed in $10^4$ stochastic simulations after pulse width optimization at both 300K and 350 K. Increasing the temperature from 300 to 350 K shifts the optimum VCMA pulse width from ~475 to ~450 ps, indicating that an apparent reduction in switching probability at a fixed pulse duration can be primarily due to pulse detuning. In contrast, strong easy-axis disorder reduces the maximum switching probability even after pulse-width optimization, indicating degradation of the switching dynamics. We further show that, under strong easy axis disorder a transient skyrmion still forms while the subsequent post-pulse dynamics fails to complete the magnetization reversal. Our study therefore establishes the distinction between recoverable reliability due to timing shifts and permanent disorder-induced degradation that simply cannot be recovered by pulse re-tuning. Finally, for no disorders but at room temperature, 100% switching is observed with $10^6$ stochastic simulations with correct pulse shaping.

## I. Introduction

Skyrmion-mediated voltage-controlled magnetic anisotropy (VCMA) is a promising switching scheme that provides low-energy (sub-fJ) and field-free magnetization reversal in perpendicular magnetic tunnel junctions (p-MTJs) [1-3]. Related skyrmion-based concepts have also been explored for voltage-controlled core reversal, multilevel memory, neuromorphic applications, and localized quantum-spin control [4-7]. A voltage pulse nucleates a skyrmion that undergoes breathing motion and the pulse must be withdrawn at the right phase during the breathing for the reversal to complete. Previous works have shown that this process is reliable under room temperature thermal noise and can also be scaled to nanoscale p-MTJs [2,3,8]. However, the switching strongly depends on pulse timing and whether it can remain reliable under material defects found in sputtered magnetic devices remains an unresolved problem.

The reliability problem is particularly important in sputtered ultrathin magnetic films, where polycrystalline structure introduces local variations in magnetic properties. Polycrystalline systems inherently have grain-to grain variation in DMI, PMA, exchange coupling and thus can pin skyrmion boundary, modify its shape and stability [9-16]. Experimental works have also shown that boundary pinning can change skyrmion stability, its motion and produce non-rigid skyrmion deformation [10-13,17]. Similarly, in VCMA switching, thermal activation and magnetic granularity can affect switching dynamics by introducing stochastic switching delays [18]. Disorders can also affect radial dynamics of a skyrmion and introduce expansion and contraction in polycrystalline structures [11,13]. Experimentally it was also shown that magnetic inhomogeneity can create pinning sites that can influence skyrmion creation and annihilation of voltage-controlled skyrmions [19-21].

As skyrmion-mediated switching is highly sensitive to pulse timing, grain-to-grain variations in DMI, PMA, easy-axis orientation, and exchange coupling can change how a skyrmion forms, breathes, and annihilates after the pulse is withdrawn. A reduction in switching probability can therefore arise for two different reasons. If disorder or temperature mainly shifts the skyrmion breathing cycle, then the reliable switching can be recovered by adjusting the pulse width. In contrast, if disorder strongly degrades skyrmion stability and its switching dynamics, then pulse-width adjustment alone may not be enough. A study that makes this distinction and identifies which type of material disorder strongly degrades and how it degrades the skyrmion-mediated switching dynamics is still lacking.

In this work, we use stochastic micromagnetic simulations to evaluate transient-skyrmion-mediated VCMA switching under grain-scale material disorder and elevated temperature. By studying switching probability over a range of pulse widths, we distinguish reliability loss caused by displacement of the favorable timing window from degradation that persists after pulse optimization. Representative trajectories are used to identify how these two regimes differ dynamically. Our results provide defect tolerance limits which can inform the design and optimization of future skyrmion-based MTJ devices and help identify when pulse width optimization is sufficient and when better control of defects is required in fabrication.

## II. Method

We simulate the magnetization dynamics of the free layer of a circular p-MTJ with a diameter of 100 nm, which is discretized into 64 × 64 × 1 cells with a cell size of 1.56 nm × 1.56 nm × 1.2 nm, well below the exchange length. The simulation uses the micromagnetic software MuMax3 [22] to solve the Landau–Lifshitz–Gilbert (LLG) equation:

$$\frac{\partial \mathbf{m}}{\partial t} = -\frac{\gamma}{1+\alpha^2}\left[\mathbf{m} \times \mathbf{B}_{\text{eff}} + \alpha\, \mathbf{m} \times (\mathbf{m} \times \mathbf{B}_{\text{eff}})\right] \quad (1)$$

where γ is the gyromagnetic ratio, α is the Gilbert damping coefficient, and m = $M/M_s$ is the normalized magnetization. $B_{eff}$ is the effective magnetic field with the following components [22,23]:

$$\mathbf{B}_{\text{eff}} = \mathbf{B}_{\text{demag}} + \mathbf{B}_{\text{exchange}} + \mathbf{B}_{\text{DMI}} + \mathbf{B}_{\text{anis}}(t) + \mathbf{B}_{\text{th}}(t) \quad (2)$$

The effective field due to perpendicular anisotropy is given by:

$$\mathbf{B}_{\text{anis}}(t) = \frac{2K_{u1}(t)}{M_s}\,(\mathbf{u}\cdot\mathbf{m})\mathbf{u} \quad (3)$$

where u is the unit vector in the anisotropy direction. VCMA-induced PMA modulation is employed by modulating $K_{u1}(t)$ while keeping $K_{u2} = 0$.

The thermal field contribution to the effective field is given by [24]:

$$\mathbf{B}_{\text{th}}(t) = \sqrt{\frac{2\alpha k_B T}{\gamma M_s V \Delta t}}\,\boldsymbol{\beta}(t) \quad (4)$$

where $k_B$ is the Boltzmann constant, T is the temperature, V is the cell volume, Δt is the time step, and η(step) is a random vector from a standard normal distribution whose value is changed after every time step.

The parameters for Co-based magnetic materials used in the simulations are given in Table 1 [1,25]. $N=10^6$ independent simulations are performed at the nominal 500-ps pulse width for the defect-free baseline, while N=$10^4$ simulations are performed for the remaining pulse-width, disorder, and temperature conditions. Switching is considered successful if the spatially averaged out-of-plane magnetization component ⟨mz⟩ < −0.8 after the post-pulse relaxation period.

To investigate the effect of material disorder on switching reliability, the 100-nm free layer is partitioned into a polycrystalline structure using Voronoi tessellation with an average grain size of 4 nm. Grain-to-grain variations in the perpendicular anisotropy $K_{u1}$ and DMI constant $D$ are introduced using a Gaussian random variable for each grain. For a disorder level $p$, the parameters are assigned as $K_{u1,i} = K_{u1,0}(1 + p\xi_i)$ and $D_i = D_0(1 + p\xi_i)$. Here $i$ denotes the grain index, $K_{u1,0}$ and $D_0$ are the nominal perpendicular-anisotropy and DMI values, respectively, given in Table 1, $p$ is the fractional DMI/PMA disorder level, and $\xi_i$ is a standard normally distributed random variable. The same $\xi_i$ is used for both $K_{u1}$ and $D$ within a given grain, so the DMI and PMA disorders are spatially co-varied. Easy-axis disorder is introduced by assigning each grain a locally tilted anisotropy axis and intergrain exchange disorder is introduced by scaling the exchange coupling between different grain regions. The disorder ranges were selected around values previously used in granular skyrmion and VCMA-MTJ simulations, including ~10% magnetic-parameter variation and 10% reduced intergrain exchange [18], while broader skyrmion-disorder studies have considered variations from a few percent to several tens of percent [14,26]. The 2–6% easy-axis disorder corresponds to angular deviations of a few degrees, comparable to experimentally reported easy-axis distributions in patterned p-MTJs [27], and the co-variation of DMI and PMA is used as a simplified correlated-disorder model motivated by experimentally observed DMI–anisotropy correlations [28].

Table 1. Parameters used in the simulation [1,25].

| **Parameters** | **Value** |
| --- | --- |
| Saturation magnetization (Ms) | $1.3\times10^{6}$ A $m^{-1}$ |
| Exchange constant (A) | $2.5\times10^{-11}$ J $m^{-1}$ |
| Perpendicular anisotropy constant ($K_{u1}$) | $1.11\times10^{6}$ J $m^{-3}$ |
| Effective perpendicular magnetic anisotropy (PMA), $K_{eff} = K_{u1} - (1/2)\ \mu_0 Ms^2$ | $0.48\times10^{5}$ J $m^{-3}$ |
| Critical DMI, $D_c = \frac{4}{\pi}\sqrt{AK_{eff}}$ | 1.42 mJ $m^{-2}$ |
| DMI constant, D | 1.2 mJ $m^{-2}$ |
| Gilbert damping (α) | 0.01 |

## III. Results and Discussion

### A. Baseline Switching Dynamics and Reliability

Figure 1 summarizes the transient-skyrmion-mediated switching process and its baseline timing dependence. Reducing $K_{u1}$ from 1.11 to 0.94 MJ $m^{-3}$ nucleates an intermediate skyrmion, which reverses the magnetization when the VCMA pulse is withdrawn during the favorable inbreathing phase. At 300 K, the reliable switching window spans 450–650 ps [Fig. 1(d)], with 100% switching observed in $10^4$ simulations at each sampled pulse width. Outside this window, pulse withdrawal does not coincide with the favorable skyrmion-breathing phase. At 500 ps, no switching failures were observed in $10^6$ simulations, demonstrating highly reliable operation at room temperature thermal noise.

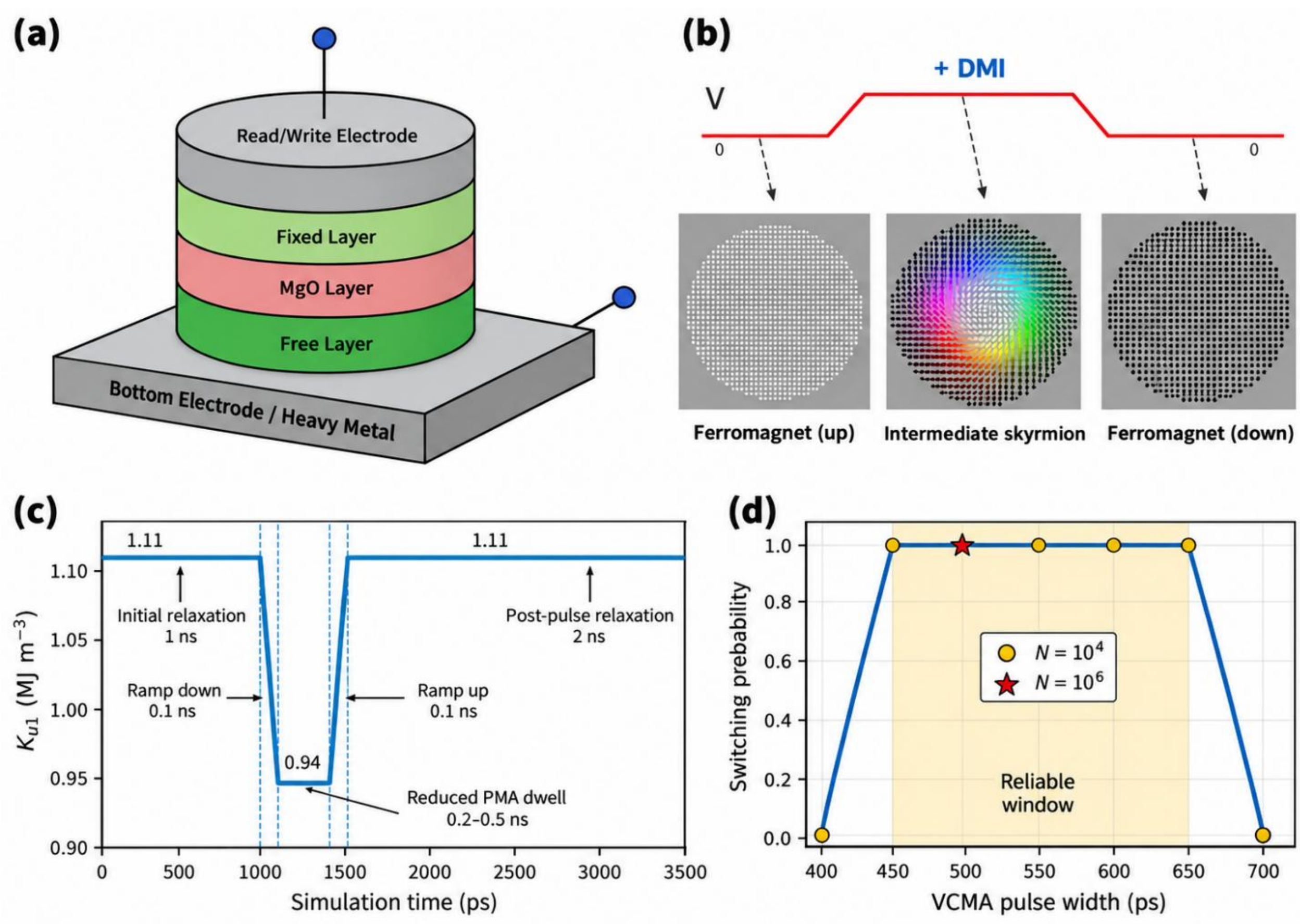

***Figure 1.*** *(a) Schematic of the p-MTJ structure. (b) Ferromagnetic up, intermediate skyrmion, and ferromagnetic down states visited during switching, shown below a schematic voltage pulse. (c) Modulation of the perpendicular anisotropy $K_{u1}$ due to the voltage pulse, showing the 0.1 ns rise and fall times and the 0.2–0.5 ns dwell duration, which depends on the pulse width. (d) Switching*

*probability versus pulse width, with N = $10^4$ independent simulations performed at each pulse width. N = $10^6$ simulations were performed at the optimal pulse width of 500 ps.*

**B. Effect of Material Disorder on Switching Reliability**

To evaluate grain-scale material tolerance, we considered spatial variations in DMI/PMA, easy-axis orientation, and intergrain exchange using the polycrystalline model described in Sec. II. Figure 2 compares the resulting switching probability at 300 K and a fixed 500-ps pulse. Switching remains highly robust up to 15% DMI/PMA variation and 2% easy-axis disorder but degrades sharply once the disorder exceeds a threshold. For 2% easy-axis misalignment, $P_{sw}$ = 100% is retained up to 15% DMI/PMA variation, whereas 6% easy-axis misalignment suppresses switching almost completely at the same DMI/PMA disorder level.

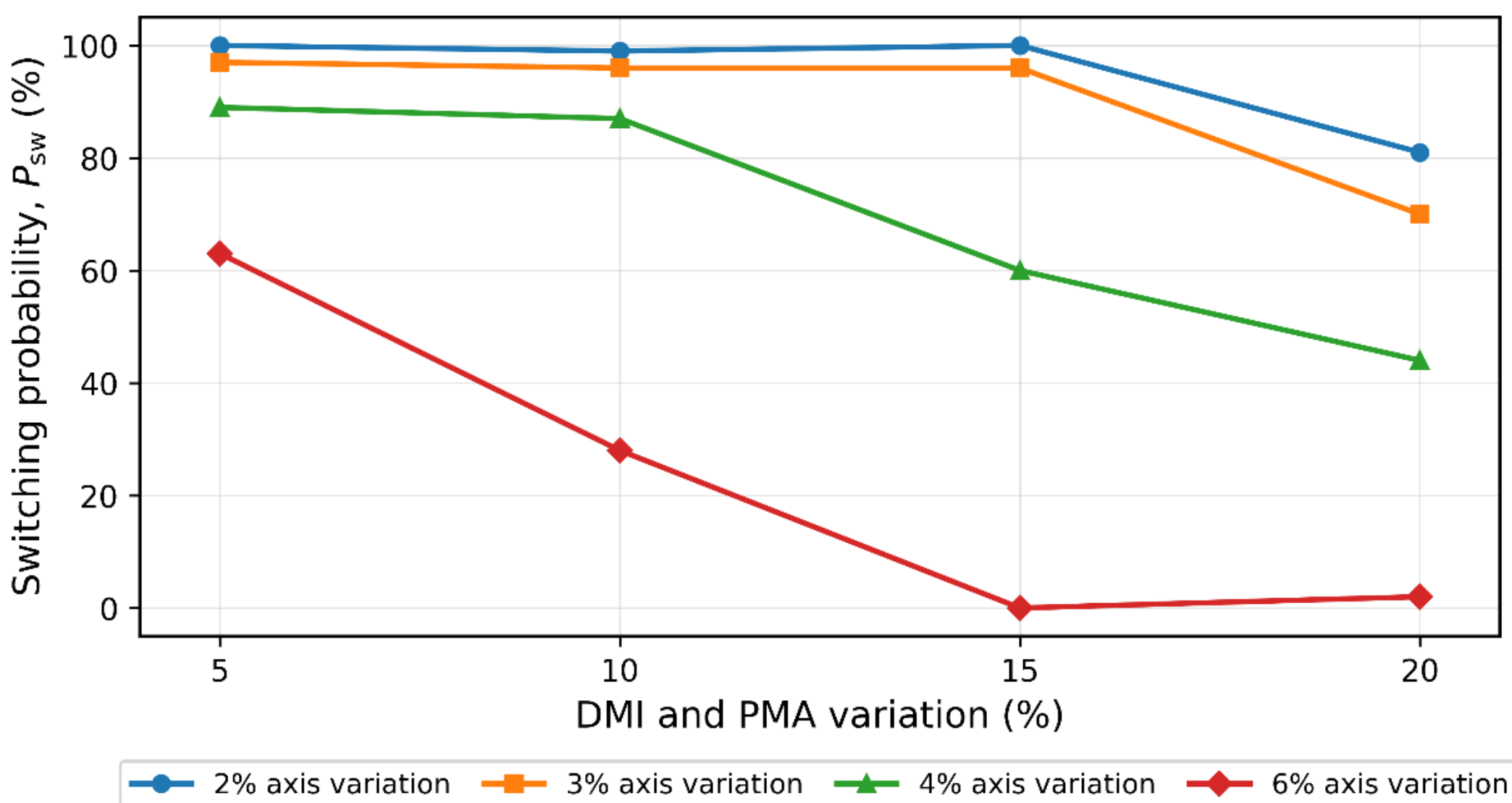

***Figure 2.*** *Switching probability under material disorder as a function of DMI and PMA variation for easy-axis misalignment of 2%, 3%, 4%, and 6%. N = $10^4$ independent simulations were performed at the fixed pulse width of 0.5 ns for each condition.*

**C. Distinguishing Pulse Detuning from Intrinsic Reliability Degradation**

Figure 3 shows switching probability ($P_{sw}$) as a function of pulse width and temperature for representative disorder conditions. The defect-free device retains a broad high-probability switching (low error) region across 300–400 K [Fig. 3(a)]. Under moderate disorder, this window shifts toward shorter pulse widths, with the optimum moving from ~475 ps at 300 K to ~450 ps at 350 K and ~425 ps at 400 K [Fig. 3(b)]. Therefore, the strong reduction in $P_{sw}$ observed at the fixed 500-ps pulse is largely recovered by pulse-width re-tuning. However, strong easy-axis disorder shows a qualitatively different response [Fig. 3(c)]. Although its optimum pulse also shifts toward shorter pulse widths, the maximum switching probability decreases with increasing temperature and remains limited to ~73% at 400 K. These findings suggest that reliability loss can arise either from a shift in the favorable pulse-width window or from degradation of the switching dynamics, which requires further investigation.

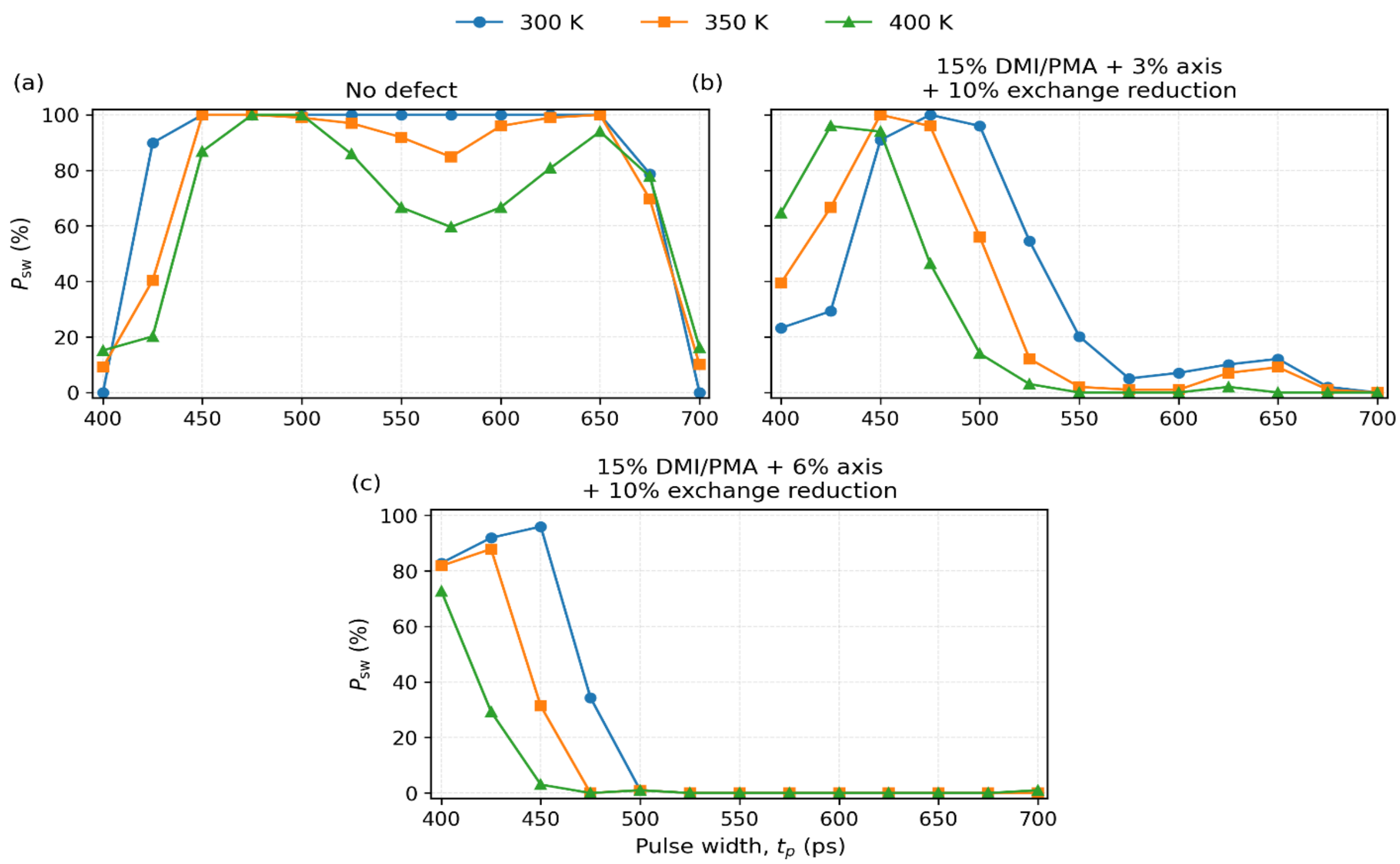

***Figure 3.*** *Switching probability as a function of VCMA pulse width and operating temperature for (a) the defect-free structure, (b) moderate disorder consisting of 15% co-varied DMI/PMA variation, 3% easy-axis misalignment, and 10% intergrain exchange reduction, and (c) strong easy-axis disorder consisting of 15% DMI/PMA variation, 6% easy-axis misalignment, and 10% intergrain exchange reduction.*

To clarify the two forms of reliability loss identified in Fig. 3, Fig. 4 compares representative switching trajectories before and after pulse-width optimization. The defect-free device switches successfully at the nominal 500-ps pulse, with formation and subsequent annihilation of the transient skyrmion followed by relaxation into the reversed state. Under moderate disorder, the 500-ps trajectory also reaches topological charge, $Q \approx 1$, but remains in that state for longer duration and fails to complete magnetization reversal. When the pulse is retuned to 425 ps, under the same disorder condition, the topological charge returns rapidly towards zero ($Q \approx 0$) and follows a successful post-pulse relaxation to reach the reversed state. However, strong easy-axis disorder behaves differently. Even at the optimized 400-ps pulse, the representative failed trajectory still forms the transient skyrmion but remains in that state for longer compared to other cases and subsequently returns toward the initial +z state instead of stabilizing in the reversed state. These trajectories indicate that strong easy-axis disorder produces a degradation of the switching dynamics by changing skyrmion stability and post-pulse relaxation dynamics that cannot be compensated by pulse retuning alone.

(a)

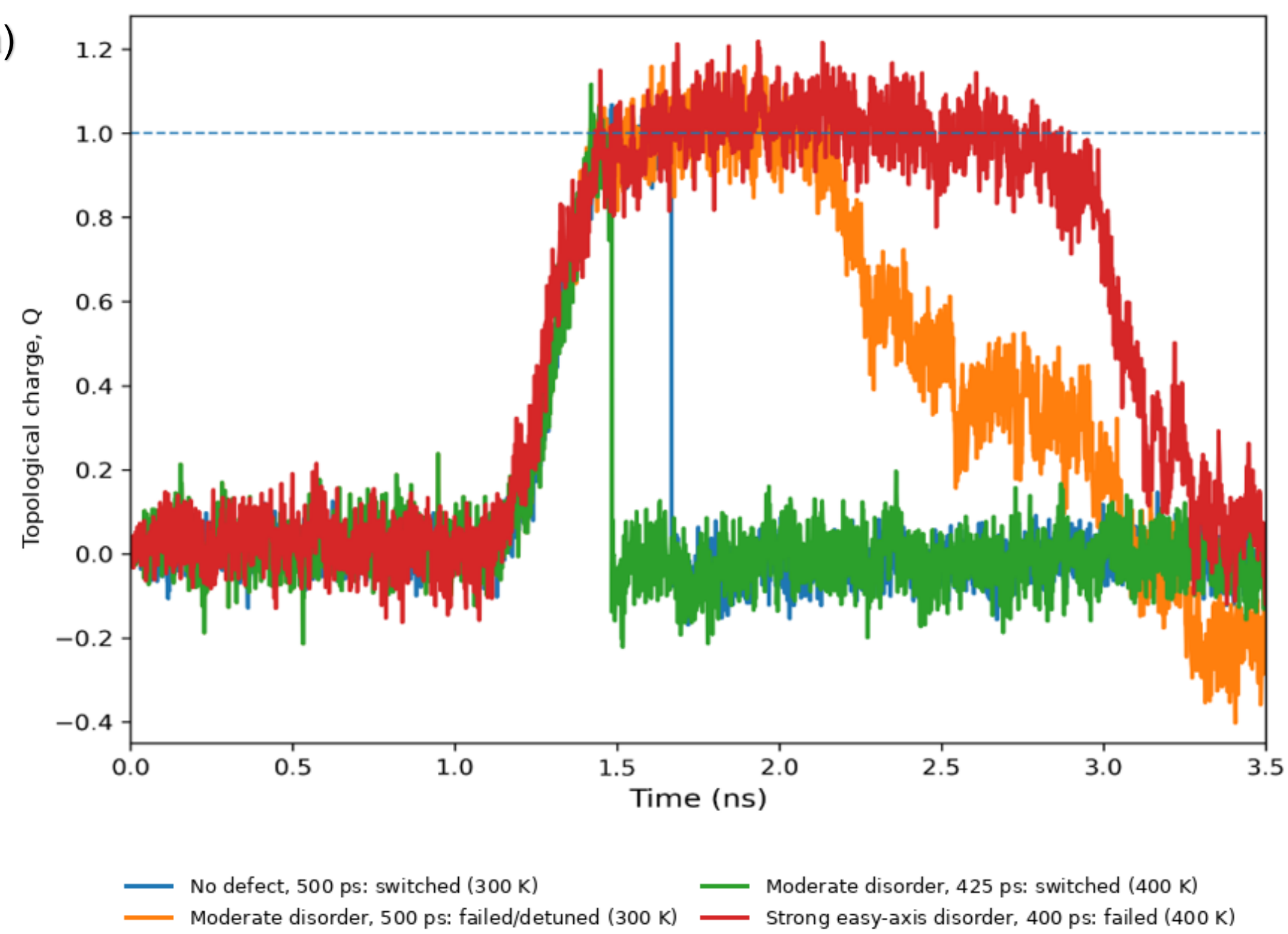

(b)

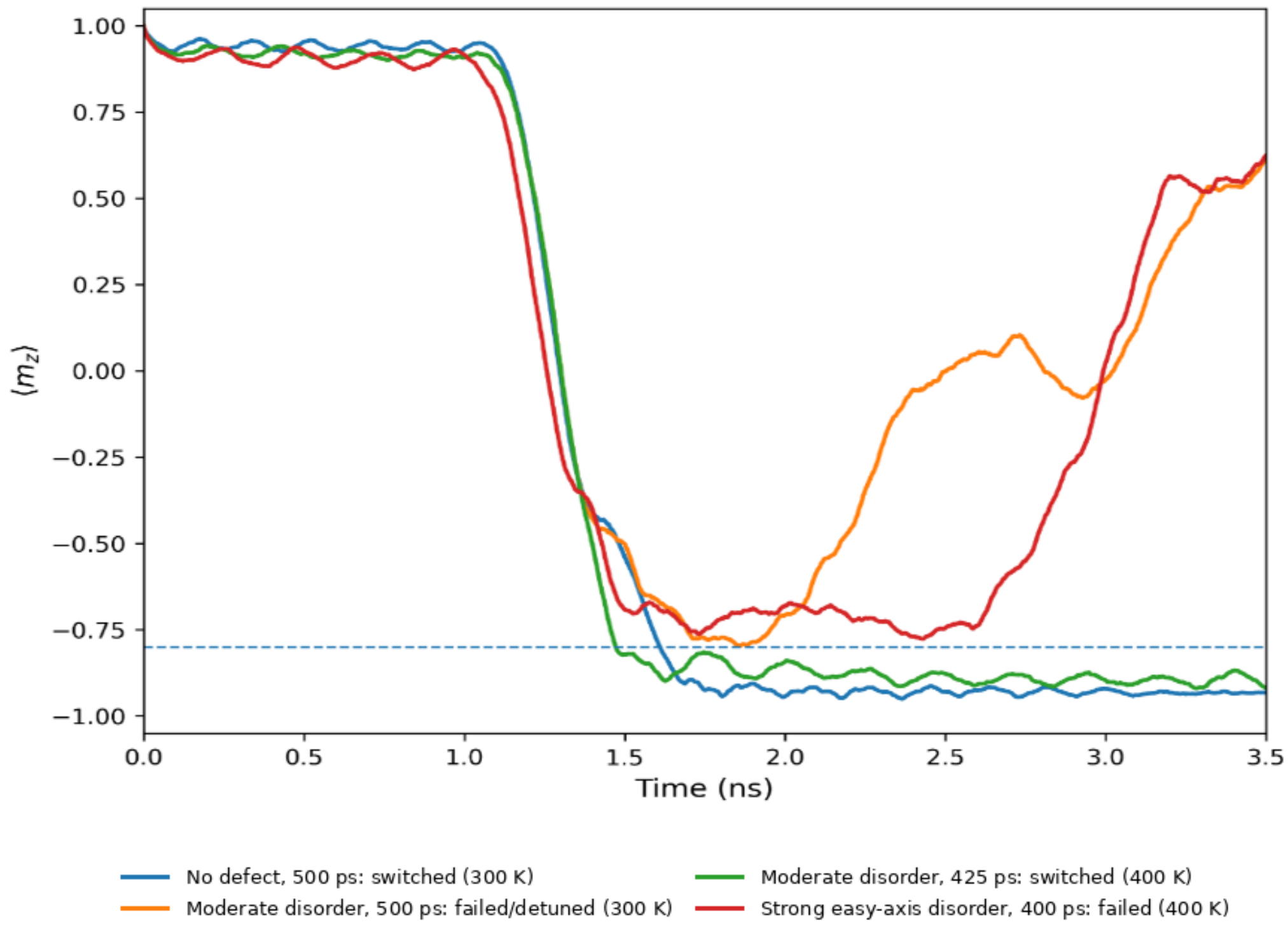

***Figure 4.*** *Representative trajectories illustrate recoverable pulse detuning and residual switching degradation. Time evolution of (a) topological charge, Q, and (b) spatially averaged out-of-plane magnetization,* $\langle m_z \rangle$*, for defect-free, moderate-disorder, and strong easy-axis-disorder cases. Pulse retuning from 500 to 425 ps restores switching under moderate disorder, whereas strong easy-axis disorder can still lead to failure at the optimized pulse width. The dashed line in (b) denotes the switching criterion,* $\langle m_z = -0.8 \rangle$*.*

Figure 5 summarizes the distinction between recoverable timing shifts and permanent reliability degradation under strong disorder. Under moderate disorder, pulse-width optimization maintains 100% switching in $10^4$ stochastic simulations at both 300 and 350 K, despite the temperature-dependent shift in the optimum pulse width. Even at 400 K, pulse retuning increases $P_{sw}$ from ~14% at the fixed 500-ps pulse to ~96%. In contrast, under strong easy-axis disorder, the maximum switching probability decreases with temperature and is limited to ~73% at 400 K even after pulse-width optimization. These results show that pulse retuning can compensate for timing detuning under moderate disorder, whereas strong disorder produces a degradation in the switching dynamics that cannot be fully recovered by pulse-width optimization.

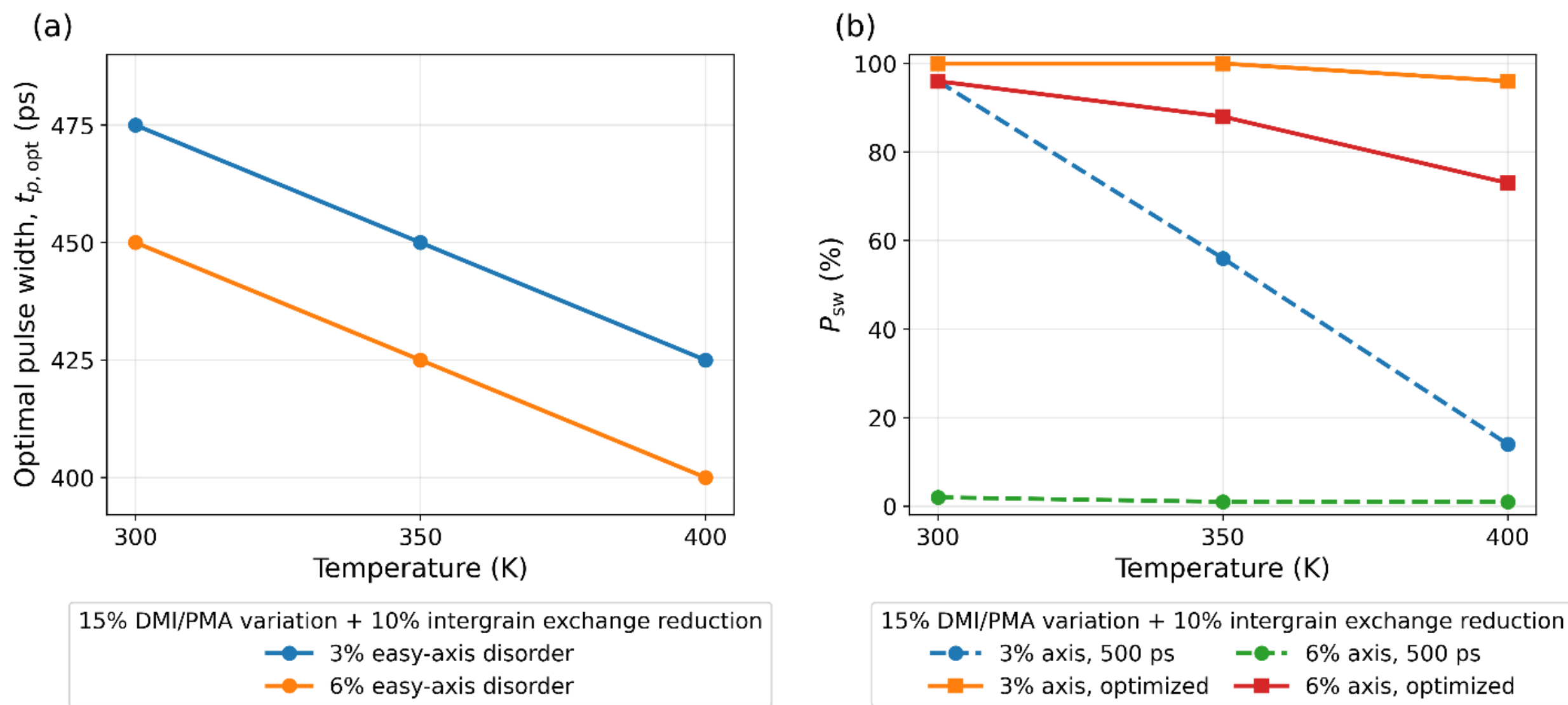

***Figure 5.*** *Effect of pulse-width optimization under material disorder and elevated temperature. (a) Optimum VCMA pulse width as a function of temperature for the three representative disorder conditions. (b) Comparison between the switching probability at the fixed 500-ps pulse and the maximum switching probability obtained by pulse-width optimization (pulse width varies with temperature and defects).*

## IV. Conclusion

In summary, we investigated how grain-scale material disorder, temperature, and pulse timing influence the reliability of transient-skyrmion-mediated VCMA switching. The results show that moderate disorder does not necessarily compromise the underlying switching mechanism. Instead, disorder and temperature can shift the transient-skyrmion dynamics and change the favorable pulse width. This explains why highly reliable switching can be maintained under moderate disorder and elevated temperature when the VCMA pulse is appropriately optimized.

A different limitation emerges when the material disorder becomes sufficiently strong. In that case, pulse retuning alone cannot fully restore switching. Our findings indicate that a transient skyrmion can still be created, but the post-pulse evolution may fail to stabilize the reversed magnetic state. The limiting factor is therefore not skyrmion nucleation, but skyrmion post-pulse stability and annihilation dynamics required for deterministic reversal.

Our findings separate two fundamentally different sources of reliability loss in skyrmion-mediated VCMA switching under material disorder: (1) pulse timing errors that can be compensated electrically and (2) disorder-induced degradation of the reversal dynamics that requires improved material uniformity. This distinction provides a practical basis for VCMA device design, where pulse optimization can be used to tolerate moderate disorder and material variations, while control of magnetic inhomogeneity sets the limit on switching reliability. These findings can help guide future simulation studies, experimental characterization, and the design of reliable, energy-efficient skyrmion-based memory devices suggesting switching errors less than $10^{-4}$ is possible even under large defects not typically encountered in large magnetic devices and at operating temperature well above room temperature. As the defects are reduced to zero, switching errors less than $10^{-6}$ appear to be feasible at room temperature.

**Acknowledgement:** Funding for this work from Virginia Commonwealth University Commercialization Fund and Virginia Innovation Partnership Corporation (VIPC) Commonwealth Commercialization Fund for Higher Education award #CCF24-0087-HE are acknowledged.

**Conflict of Interest Disclosure:** Although no part of this work was funded by Rocket One, Inc. Dr. Jayasimha Atulasimha is an inventor of intellectual property related to this project, which is licensed by Rocket One, Inc. Dr. Jayasimha Atulasimha also receives payments from Rocket One, Inc. for external consulting with the company.